\documentclass{article}
\usepackage{spconf,amsmath,graphicx}
\usepackage{hyperref}
\usepackage{enumitem}
\setlist{nosep, leftmargin=14pt}

\usepackage{mwe} 

\title{Learning with less labels in Digital Pathology\\via Scribble Supervision from Natural Images}
\name{Eu Wern Teh and Graham W.~Taylor}
\address{School of Engineering, University of Guelph, ON, Canada\\
    Vector Institute, ON, Canada}
\begin{document}
%
\maketitle
\begin{abstract}


A critical challenge of training deep learning models in the Digital Pathology (DP) domain is the high annotation cost by medical experts. One way to tackle this issue is via transfer learning from the natural image domain (NI), where the annotation cost is considerably cheaper.
Cross-domain transfer learning from NI to DP is shown to be successful via class labels~\cite{teh2020learning}. One potential weakness of relying on class labels is the lack of spatial information, which can be obtained from spatial labels such as full pixel-wise segmentation labels and scribble labels. We demonstrate that scribble labels from NI domain can boost the performance of DP models on two cancer classification datasets (Patch Camelyon Breast Cancer and Colorectal Cancer dataset). Furthermore, we show that models trained with scribble labels yield the same performance boost as full pixel-wise segmentation labels despite being significantly easier and faster to collect.



\end{abstract}
\begin{keywords}
Transfer Learning, Annotation Efficient Learning
\end{keywords}

\begin{figure}[htb]
  
  \centering
  \includegraphics[width=2.80in,   ]{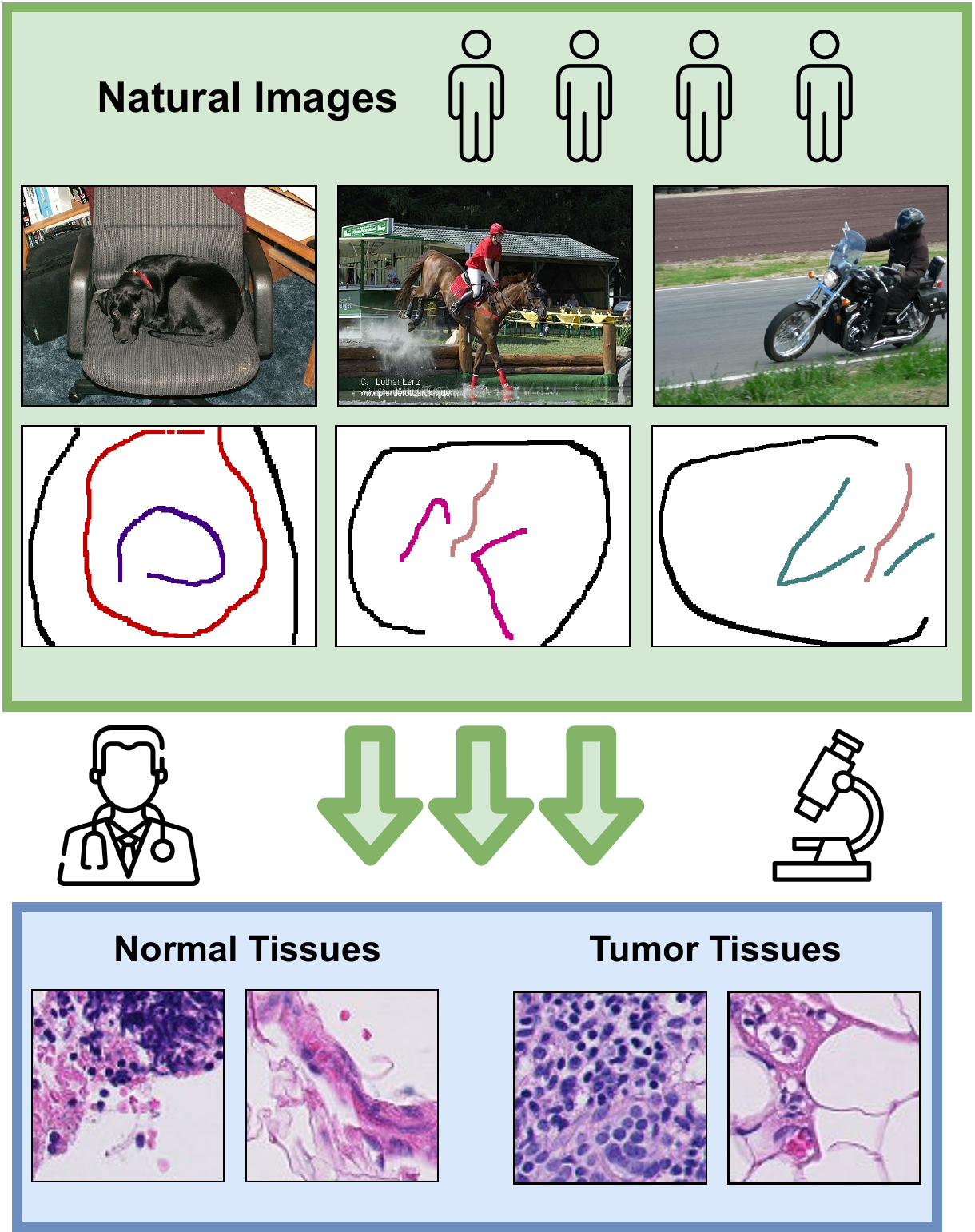}
  \caption{
  A demonstration of our proposed approach in cross-domain transfer learning. We first train a model by using scribble labels from the natural image (NI) domain. We transfer knowledge from the NI domain to the Digital Pathology (DP) domain by initializing the DP models with the pre-trained weights. Lastly, we train these DP models using labels provided by medical experts on cancer classification tasks.
  }
  \label{fig:main}
\end{figure}

\section{Introduction}\label{sec:intro}

Digital Pathology (DP) is a field that involves the analysis of microscopic images.
Supervised learning has made progress in DP for cancer classification and segmentation tasks in recent years; however, it requires a large number of labels to be effective~\cite{srinidhi2020deep}.
Unfortunately, labels from medical experts are scarce and extremely costly. On the other hand, it is relatively cheap to obtain labels from a layperson in the natural image (NI) domain.
We leverage inexpensive annotations in the NI domain to help models to perform better in DP tasks (Fig~\ref{fig:main}). 
We also investigate the effectiveness of spatial NI labels as they vary by level of detail and thus annotator effort.

Transfer learning via pre-training allows a model to learn features from a source task and transfer them to a target task to improve generalization.~\cite{huh2016makes}.
Features learned from image class labels in the NI domain are shown to be useful in boosting the performance of DP models, especially in the low-data regime~\cite{teh2020learning,kupferschmidt2021strength}.
However, class labels are devoid of spatial information, which can be crucial in transfer learning, and this information can be obtained from spatial labels (e.g., full pixel-wise segmentation labels and scribble labels).
In terms of the level of detail in labels, scribble labels are far weaker than full pixel-wise segmentation labels. However, collecting full pixel-wise segmentation labels is extremely laborious and time-consuming, making it less attractive than scribble labels. Despite the lack of details in scribble labels, we discover that models trained with scribble labels from the NI domain perform equally well in transfer learning compared to models trained with full pixel-wise segmentation labels. This discovery allows us to reliably use scribble annotations, which is significantly faster to obtain than the collection of full pixel-wise segmentation labels.


\section{Related Work}\label{sec:related}

Deep Learning models are effective in a fully supervised environment, but they require tremendous amounts of human labels~\cite{goodfellow2016deep}. However, human labels are often scarce in the Digital Pathology (DP) domain due to high annotation costs~\cite{srinidhi2020deep}.
Transfer learning in the form of pre-training is an effective way to combat label scarcity~\cite{huh2016makes}. Models are commonly pre-trained with class labels from natural image (NI) datasets~\cite{imagenet_cvpr09,mahajan2018exploring}, and these pre-trained models have helped to boost the performance of many computer vision tasks~\cite{huh2016makes}.

Conventionally, spatial labels are obtained to solve spatial tasks such as semantic segmentation~\cite{chen2017deeplab}, object detection~\cite{Redmon_2017_CVPR}, and human pose estimation~\cite{he2017mask}. Among all spatial labels, collecting full pixel-wise segmentation labels is considered as one of the most time-consuming tasks~\cite{lin2014microsoft}.
Scribble labels were introduced to reduce the reliance on full pixel-wise segmentation~\cite{lin2016scribblesup}. The collection of scribble labels is significantly faster than full pixel-wise segmentation labels. On a low-data regime, joint training of scribble labels and full pixel-wise segmentation labels are shown to be better than models trained only on full pixel-wise segmentation labels. Other works have shown that optimizing graph models can further improve scribble-aided segmentation models~\cite{tang2018regularized,tang2018normalized}.




\section{Methods}\label{sec:methold}

We use a 2D Cross-Entropy Loss as described in Equations~\ref{eq:1} and~\ref{eq:2} to train our models using the full pixel-wise segmentation labels and the scribble labels. Both equations describe the loss for a single image, $x$, and the corresponding spatial mask, $y$, each of dimension $I \times J$, $y_{i,j} \in \{0,1,2,...K\}$. The function, $f$ denotes the segmentation model, $K$ indicates the number of classes present during training, and 0 represents the ignore label.
\begin{align}
L_{i,j}  =  -\frac{1}{I*J}\sum_{i}^{I}\sum_{j}^{J}\log\frac{\text{exp}(f(x_{i,j})[y_{i,j}])}{\sum_{k=1}^{K}\text{exp}(f(x_{i,j})[k])}\label{eq:1}
\end{align}

\begin{figure}[htb]
  \centering
  \includegraphics[width=3.3in, ]{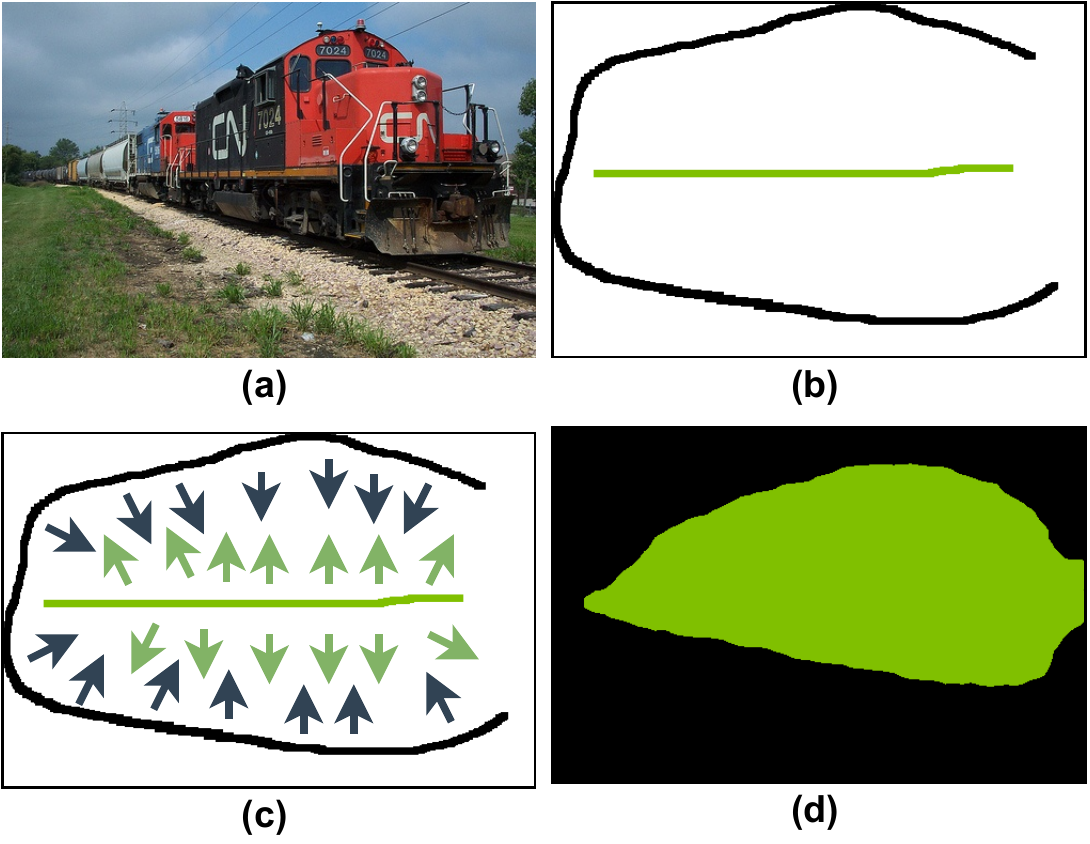}
  \caption{An illustration of how scribble supervised segmentation works. Panels (a) and (b) show the input image and the corresponding scribble mask. Panel (c) depicts the expansion direction of the model's pixel prediction. Panel (d) shows the model's pixel prediction after training.
  }
  \label{fig:voc_scrib}
\end{figure}

\begin{align}\label{eq:2}
        L_{i,j}^{'}= \begin{cases}
          L_{i,j}, & \text{if}\; y_{i,j} \neq 0,\\
            0, & \text{if}\; y_{i,j} = 0,
        \end{cases}
\end{align}

The ignore-label (white region in Fig.~\ref{fig:voc_scrib}(b)) plays an important role in scribble-supervised segmentation training.
If we treat the white region as a separate class, the model will not expand its prediction of the real classes. By incurring zero loss in the ignore-label region, the model is free to predict anything. During training, the model's predictions tend to start near the scribble annotation, and they gradually expand outward as training progresses. In Fig.~\ref{fig:voc_scrib}(c), we see that both the background and locomotive classes attempt to fill up the empty ignore-labels region. After training, an equilibrium is reached, and the final prediction is shown in Fig.\ref{fig:voc_scrib}~(d).

\section{Experiments}\label{sec:exp}

We hypothesize that models pre-trained on scribble labels can help DP models achieve higher performance when compared to models pre-trained on class labels. We validate our hypothesis by using one natural image dataset and two digital pathology datasets. We first pre-train our models on the natural image dataset before training on the digital pathology datasets. We report our results by using accuracy as the performance metric on the digital pathology datasets.

\begin{figure}[htb]
  \centering
  \includegraphics[width=3.00in, ]{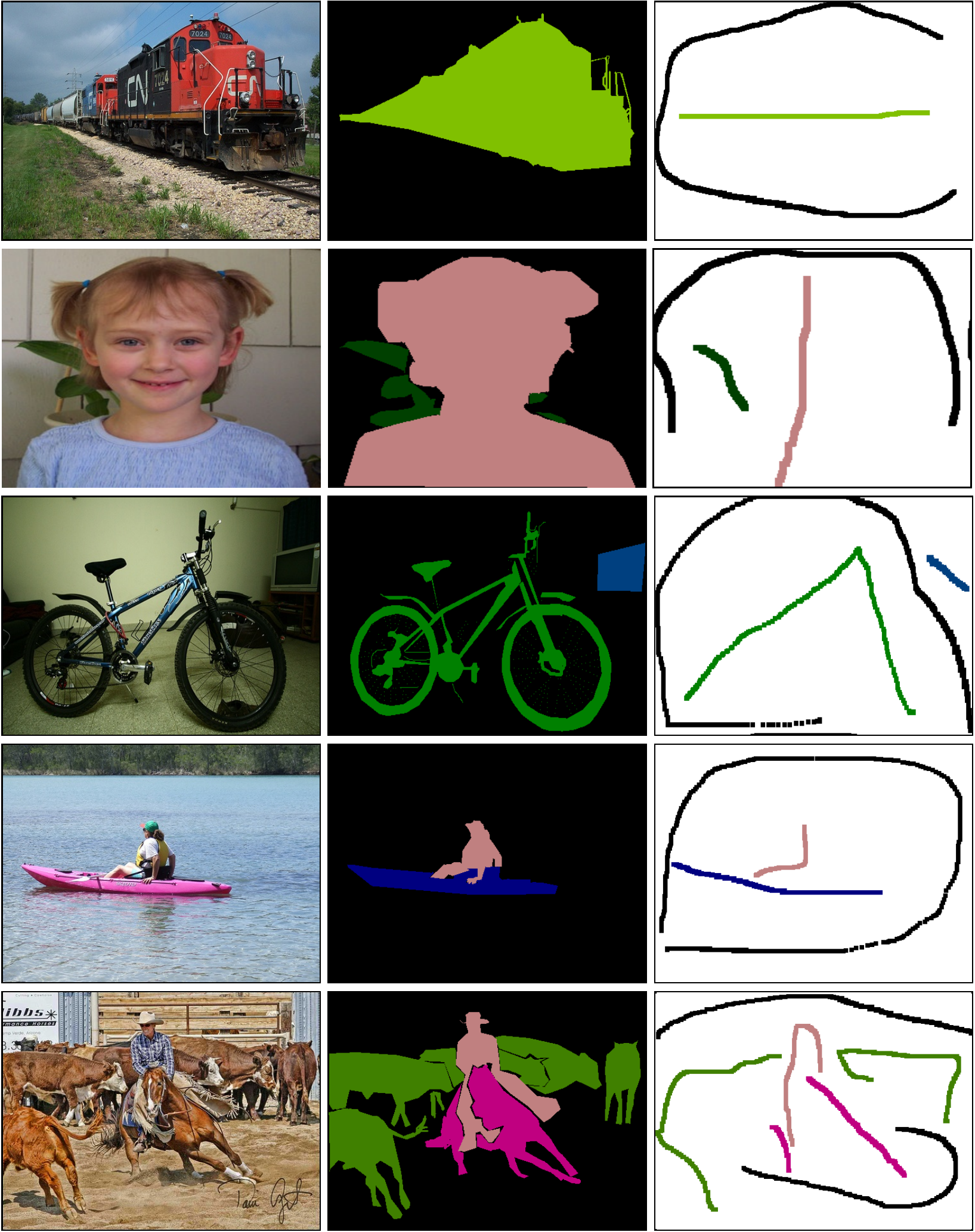}
  \caption{Examples of full pixel-wise segmentation masks and the corresponding scribble masks.
  }
  \label{fig:voc_scrib_sample}
\end{figure}

\subsection{Pre-training on Pascal VOC 2012 dataset}

We use the Pascal VOC 2012 dataset as our natural image dataset~\cite{pascal-voc-2012}. It consists of 21 classes: aeroplane, bicycle, bird, boat, bottle, bus, car, cat, chair, cow, table, dog, horse, motorcycle, person, plant, sheep, couch, locomotive, television, and background (Fig.~\ref{fig:voc_scrib_sample}). It has a total of 10,582 training images and 1,449 validation images. Additionally, we add scribble labels obtained from ~\cite{lin2016scribblesup} to our experiments.

We pre-train our models using three different settings: multi-label classification, full pixel-wise segmentation, and scribble segmentation.
We use a randomly initialized ResNet-34 model to train the multi-label classification model and
a randomly initialized DeepLabV2 model with ResNet-34 backbone for the full pixel-wise segmentation and scribble segmentation pre-training~\cite{he2016deep,chen2017deeplab}. We use the binary cross-entropy loss for multi-label classification training and the 2D cross-entropy loss for segmentation training.  We train all models for 40,000 iterations using Stochastic Gradient Descent with a learning rate of 2.5e-4, a momentum of 0.9, and a weight decay of 5e-4. We also use a polynomial decay rate of $\lambda(1-\frac{\text{iter}}{\text{max\_iter}})^{0.9}$, where $\lambda$ is the base learning rate~\cite{chen2017deeplab}. We augment the dataset via random cropping ($321\times321$), random horizontal flipping (0.5), and random re-scaling (0.5 to 1.5).
After pre-training, we initialize the downstream models with the corresponding ResNet-34's weights.


\subsection{Training and Evaluation on Digital Pathology downstream tasks}

We evaluate transfer learning performance using two downstream datasets: the Patch Camelyon and the Colorectal cancer dataset following~\cite{teh2020learning,kupferschmidt2021strength}. After pre-training a model on the Pascal VOC 2012 dataset, we initialized the model with the corresponding pre-trained weights and trained on the downstream dataset.
Following \cite{teh2020learning}, we train our downstream tasks with the Adam optimizer with a learning rate of 1e-4 using the exponential learning rate decay scheduler with a factor of 0.94. We train all our models on the PCam dataset for 100 epochs and 200 epochs for the CRC dataset. All downstream training is repeated ten times with random seeds.
Finally, we evaluate the model on the test set of the downstream dataset by using accuracy as a performance metric. The mean accuracy and standard deviation are reported in Table~\ref{table:exp_colon_pcam}.

\begin{figure}[htb]
  \centering
  \includegraphics[width=3.30in, ]{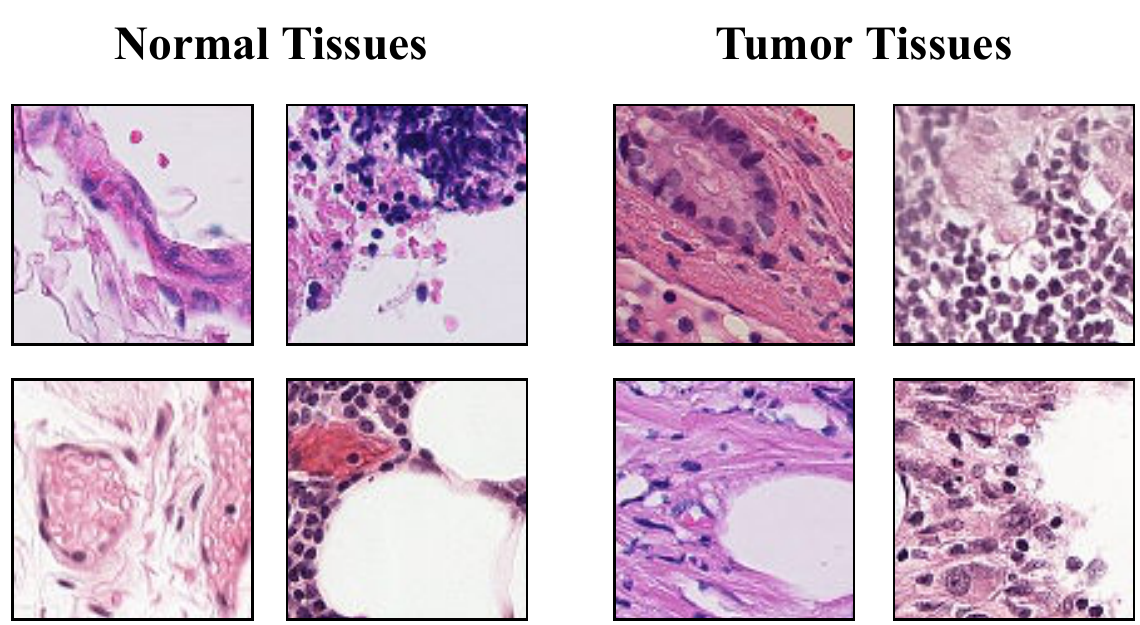}
  \caption{Tissue samples from the Patch Camelyon dataset.}
  \label{fig:pcam}
\end{figure}

The Patch Camelyon dataset (PCam) consists of 262,144 training images and 32,678 test images~\cite{veeling2018rotation}. It is a subset of the CAMELYON16 breast cancer dataset. Each patch has a resolution of 0.97 $\mu$m and a dimension of 96$\times$96 pixels. There are two classes present in this dataset: normal and tumor tissues (Fig.~\ref{fig:pcam}).

\begin{figure}[htb]
  \centering
  \includegraphics[width=3.30in, ]{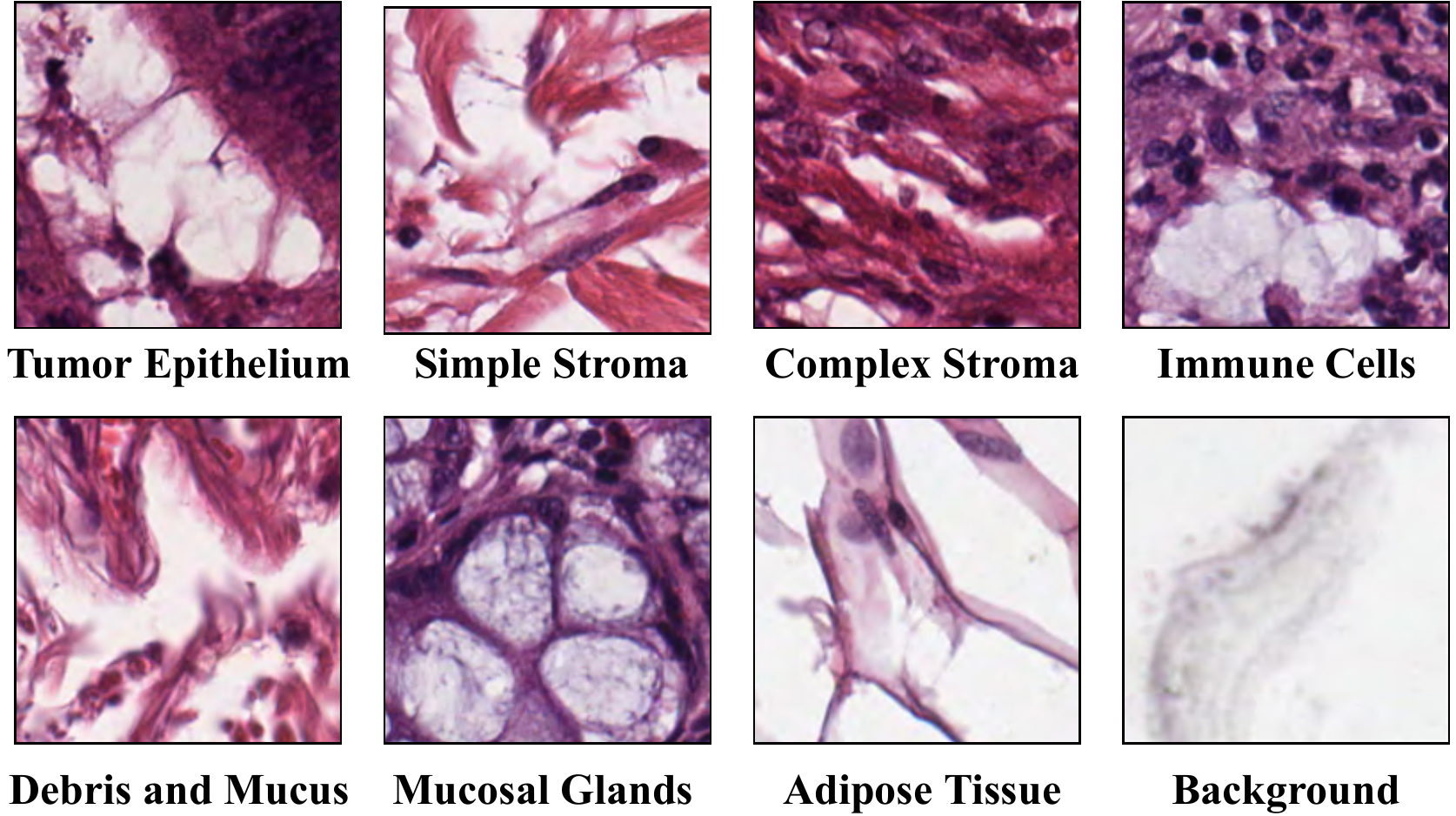}
  \caption{Tissue samples from the Colorectal cancer dataset.}
  \label{fig:crc}
\end{figure}

\begin{table*}[htb]
\centering
\setlength{\tabcolsep}{10pt}
\begin{tabular}{|l|ccc|ccc|} \hline
&\multicolumn{3}{c|}{CRC dataset}&\multicolumn{3}{c|}{PCam dataset} \\ \hline
 Initialization/Method & $N_c$ & $R$\%  & Accuracy (\%)& $N_c$ & $R$\% & Accuracy (\%) \\ \hline
Random &12 & 2  & 61.62 $\pm$ 3.79 & 1,000 & 0.76 & 79.37 $\pm$ 1.33 \\
Classification &  &   & 62.98 $\pm$ 3.70  &  &   & 76.61 $\pm$ 1.83\\
Full Segmentation &  &   & 67.34 $\pm$ 3.80 & &  & 82.11 $\pm$ 1.28\\
Scribble Segmentation &  &   & \textbf{67.40 $\pm$ 3.71} & &  & \textbf{82.13 $\pm$ 1.49}\\\hline
Random & 25 & 4 &  64.12 $\pm$ 3.73 & 2,000 & 1.53  & 81.26 $\pm$ 1.47\\
Classification & &  &  69.16 $\pm$ 3.87 &   &   & 80.95 $\pm$ 1.18\\
Full Segmentation & &  & \textbf{73.34 $\pm$ 2.17}&  &  & \textbf{86.11 $\pm$ 1.13}\\
Scribble Segmentation & &  & 72.60 $\pm$ 2.82 &  &  & 86.04 $\pm$ 1.08\\\hline
Random & 50 & 9 &  77.34 $\pm$ 2.11 & 3,000 & 2.29  & 84.09 $\pm$ 1.24\\
Classification & &  &  78.00 $\pm$ 1.25 &   &   & 82.50 $\pm$ 1.10\\
Full Segmentation & &  &  \textbf{81.40 $\pm$ 2.05} & &   & \textbf{87.80 $\pm$ 0.67}\\
Scribble Segmentation & &  &  81.02 $\pm$ 2.57 & &   & 87.68 $\pm$ 0.64\\\hline
\end{tabular}
\bigskip
    \caption{Accuracy of our model trained with $R$\% of data in four different pre-trained settings on the CRC dataset and PCam dataset. $N_c$ denotes the number of examples per class. Models for classification and segmentation are pre-trained on the Pascal VOC 2012 dataset~\cite{pascal-voc-2012}. }
    \label{table:exp_colon_pcam}
\end{table*}

The Colorectal cancer (CRC) dataset consists of 5,000 images~\cite{kather2016multi}. Each patch has a resolution of 0.49 $\mu$m and a dimension of 150$\times$150 pixels. It consists of eight classes: tumor epithelium, simple stroma, complex stroma, immune cells, debris and mucus, mucosal glands, adipose tissue, and background (Fig.~\ref{fig:crc}). We follow the same experimental setup as~\cite{teh2020learning}, where we train and evaluate our models with 10-fold cross-validation.

\subsection{Discussion}

Table~\ref{table:exp_colon_pcam} reveals that models pre-trained on spatial labels outperform the randomly initialized models and those pre-trained on Pascal class labels. On average, models pre-trained on spatial labels outperform class labels by $3.5\%$ in the Colorectal cancer dataset and  $5.3\%$ in the Patch Camelyon dataset.
These results show that spatial information matters in transfer learning from the NI to DP domain.

Models pre-trained on scribble labels perform equally well when compared to the models pre-trained on full pixel-wise segmentation labels. The average difference of mean and standard deviation between these two settings are 0.23$\%$ and 1.95$\%$ respectively. These results indicate that we can rely on scribble labels for cross-domain transfer learning.

Additionally, models pre-trained on Pascal \emph{class labels} slightly outperform the random baseline on the Colorectal cancer dataset, but they perform poorly on the Patch Camelyon dataset.
We speculate that it takes more images ($>10K$) and more than diversity ($>20$ classes) in the class labels to be useful for transfer learning.


\section{Conclusion}\label{sec:con}

We demonstrate that spatial information matters in transfer learning from the natural image (NI) domain to the Digital Pathology (DP) domain. Surprisingly, on both the CRC and PCam datasets, we showed that full pixel-wise segmentation labels are not needed for pre-training a model on the NI domain, and models pre-trained on scribble labels are sufficient to ensure a performance boost in the DP domain.

\bibliographystyle{IEEEbib}
\bibliography{isbi}

\end{document}